\documentstyle[12pt,psfig,epsf]{article}
\newcommand{\NN}{{\bf N}}
\newcommand{\ZZ}{{\bf Z}}
\newcommand{\TV}{{\cal T}}

\newcommand{\Aa}{{\cal A}}

\begin{document}

\title{Simultaneous quantization of edge and bulk Hall conductivity}
\author{Hermann Schulz-Baldes, Johannes Kellendonk, Thomas Richter 
\\
\
TU Berlin, FB Mathematik,
Stra{\ss}e des 17. Juni 136, 10623 Berlin, Germany
}

\date{}

\maketitle

\begin{abstract}

The edge Hall conductivity is shown to be an integer multiple of
$e^2/h$ which is almost surely independent of the choice of the
disordered configuration.  Its equality to the bulk Hall conductivity
given by the Kubo-Chern formula follows from K-theoretic
arguments. This leads to quantization of the Hall conductance for any
redistribution of the current in the sample.  It is argued that in
experiments at most a few percent of the total current can be carried
by edge states.

\end{abstract}


%
%

\vspace{.5cm}

Soon after the discovery of the integer quantum Hall effect (QHE)
\cite{PG}, several geometric interpretations of the observed
quantization of the Hall conductance of a two dimensional electron gas
were put forward in the framework of non-relativistic quantum
mechanics. Laughlin proposed an adiabatic Gedankenexperiment in order
to calculate the Hall conductance \cite{Lau}, Halperin and later on
B\"uttiker studied the conduction by edge channels \cite{Hal,But},
while Thouless, Kohmoto, Nightingale and den Nijs investigated the
Hall conductivity as given by the  Kubo formula \cite{TKN}.
Laughlin's argument was rigorously analyzed by Avron, Seiler and Simon
even for multiparticle Hamiltonians and in presence of a  disordered
potential \cite{ASS1,AS,ASS}.  Bellissard, recently joint by van Elst
and Schulz-Baldes, generalized the TKN$_2$-work in order to show
quantization of the Hall conductivity also in presence of a disordered
potential as long as the Fermi level lies in a region of dynamically
localized states \cite{Bel,BES}, a result that was also obtained by
Aizenman and Graf \cite{AG}. All these beautiful mathematical
approaches exhibit the Hall conductance and conductivity respectively
to have a deep geometrical meaning and allow to calculate them as an
index of a certain Fredholm operator. In \cite{TKN,ASS,BES,AG}, the 
edges of the sample play no particular r\^ole.  

\vspace{.2cm}

Recently there has been a revived interest  in edge states of magnetic
Schr\"odinger operators.  Hatsugai linked an edge state winding number
to the Chern numbers for Harper's equation with rational flux
\cite{Hat}.   Akkermans, Avron, Narevich and Seiler introduced
spectral boundary conditions giving rise to a linear dispersion
relation for edge states and a natural setting for the  Laughlin wave
function as a many body bulk state \cite{AANS}.  The stability of the
absolutely continuous spectrum associated to edge states under the
perturbation with a random potential was studied by several authors
with Mourre's positive commutator estimates \cite{MMP,BP,FGW}.  

\vspace{.2cm}

Our first main result is a rigorous proof of the edge current
quantization in the sense of Halperin for a discrete magnetic
half-plane operator containing a disordered potential, notably we show
quantization of what we call the edge Hall conductivity. Our second
mathematical result is its equality to the bulk Hall conductivity as
calculated by the Kubo-Chern formula \cite{TKN,Bel,BES}. The proof of
this equality unveals a deep connection between the plane and edge
geometry as it is based on Bott periodicity, the heart of K-theory
\cite{Bla}. We still need a gap in the spectrum of the plane operator,
but a generalization to a region of dynamically localized states is under
investigation. Using these results, we reproduce Halperin's argument
explaining why the Hall conductance of a Hall bar is quantized no
matter what proportion of the current is actually carried by the edge
or the bulk states respectively. Finally we present a simple
theoretical reasoning showing that in a typical experimental situation
at most 10\% of the current flows by edge states.

\vspace{.2cm}

For the definition of the edge Hall conductivity, we consider a gas of
charged independent particles in the (discrete) upper half plane
$\Gamma=\{(x,y)\in\ZZ^2|y\geq 0\}$ submitted to a perpendicular
magnetic field $B$.  Let $\hat{H}$ denote the one-particle Hamiltonian
acting on $\ell^2(\Gamma)$.  All operators on the half-plane space
carry a hat from now on.  Typically $\hat{H}$ is the projection onto
$\ell^2(\Gamma)$ of an operator $H=H_H+V$ acting on $\ell^2(\ZZ^2)$
where $H_H$ is Harper's magnetic Hamiltonian and $V$ is the sum of a
periodic and a disordered potential.  As the edge of the plane
intercepts the cyclotron orbits, there may be a net electric current
along the edge.  In order to calculate it, let $J$ be a spectral
interval lying in a gap of the plane Hamiltonian $H$.  Such an
interval typically contains extended edge states of $\hat{H}$
\cite{Hat}, even in presence of a weak disordered potential
\cite{Hal,MMP,BP,FGW}.  If $\hat{P}_J$ is the spectral projection of
$\hat{H}$ on $J$, then the electric edge current in  $x$-direction
carried by the eigenstates in $J$ is equal to $q\hat{{\cal
T}}(\hat{P}_J\nabla_x(\hat{H}))/\hbar$. Here $q$ is the charge of the
particles,  $\nabla_x(\hat{H})=\imath [X,\hat{H}]$ is the current
operator given by the commutator of the Hamiltonian and the
$X$-position operator and finally the trace $\hat{{\cal T}}=\mbox{Tr}_y
{\cal T}_x$ is the  trace per unit volume \cite{Bel,BES} in the
$x$-direction and the usual trace in $y$-direction. Now given an
energy $E$ in a gap of extended states of $H$, we define 

\begin{equation}
\label{eq-edgecond}
\sigma^e_{\perp}(E)
\;=\;
\frac{q^2}{\hbar}
\lim_{J\to \{E\}}\,\frac{1}{|J|}\,
\hat{{\cal T}}(\hat{P}_J\nabla_x(\hat{H}))
\mbox{ . }
\end{equation}

\noindent Because an infinite half-plane is a typical model for a
mesoscopic volume with a boundary, we call $\sigma^e_{\perp}(E)$ the
edge Hall conductivity rather than the edge Hall conductance just as
the bulk Hall conductivity is calculated with an infinite planar
model for a mesoscopic volume, while the conductance is always
associated to a finite macroscopic sample. Both the edge and the bulk
Hall conductivity are idealized quantities for which clear
mathematical results can be obtained. Further we note that one could
define the edge Hall conductivity for a strip geometry, but this would
not lead to quantization because of backscattering, that is tunneling
from upper to lower edge states \cite{But}.

\vspace{.2cm}

Before starting the more mathematical analysis, let us consider the
Harper Hamiltonian $H_H$ on $\ell^2(\ZZ^2)$  in order to familiarize
ourselves with the notion of edge Hall conductivity.  It is defined by
the finite difference equation
$(H_H\psi)_{n,m}=\psi_{n+1,m}+\psi_{n-1,m}+ e^{2\pi\imath
\varphi}\psi_{n,m+1}+ e^{-2\pi\imath \varphi}\psi_{n,m-1}$ and we
suppose   here that the magnetic flux per unit cell is rational
$\varphi=p/q$.  Then  the spectrum of $H_H$ is known to be a band
spectrum.  To analyse the half plane operator $\hat{H}_H$ on
$\ell^2(\Gamma)$, we use the translation invariance in the
$x$-direction to make a Bloch decomposition $\hat{H}_H=
\int^\oplus_{\bf [-\pi,\pi)}\frac{dk_x}{2\pi} \,\hat{H}_H(k_x)$ where
$\hat{H}_H(k_x)$  is a Jacobi matrix on $\ell^2(\NN)$.  The spectrum
of each $\hat{H}_H(k_x)$ contains the bands of the corresponding
periodic  operator $H_H(k_x)$ on $\ell^2(\ZZ)$, but there may now also
be a Dirichlet eigenvalue $\hat{E}_l(k_x)$ in each gap of
$H_H(k_x)$ \cite{Hat}. 
Upon varying $k_x$, the eigenvalues form a finite number
of continuous curves the end-points of which touch the adjacent Bloch
bands of $H_H$ (see Fig. 1). To each of these so-called {\sl edge
channels} we associate a weight $+1$ (respectively $-1$) if the
Dirichlet eigenvalues of the channel vary from the upper towards the
lower (respectively lower to upper) adjacent Bloch band as $k_x$
increases. Let $s_n$ be the sum of all these weights in the $n$th gap
$G_n$ of $H_H$.  Then the edge current carried by the edge states in
an interval $J$ contained in $G_n$ is equal to $s_n|J|q^2/h$ because
$$
\hat{{\cal T}}(\hat{P}_J\nabla_x(\hat{H}_L))
\;=\;
\sum_{l} \int_{-\pi}^{\pi} dk_x \,
\chi_J(\hat{E}_l(k_x))\frac{d\hat{E}_l(k_x)}{dk_x}
\mbox{ . }
$$ 
Here $\chi_J$ denotes the indicator function on $J$.  This implies
that $\sigma^e_\perp(E)=s_nq^2/h$  for all $E\in G_n$. Hatsugai, in a
beautiful paper \cite{Hat}, has shown that $s_n$ is equal to the sum
of the Chern numbers of the $n$ bands below $G_n$. This sum multiplied
by $q^2/h$ is the bulk Hall conductivity $\sigma_\perp^b(E)$
\cite{TKN}. Hence we obtain $\sigma_\perp^e(E)=\sigma_\perp^b(E)$ for
all energies in the gaps of $H_H$, which is a particular case of
Theorem 2 below.  Note that the equivalent result for the Landau
Hamiltonian  simply states that there are $n$ edge channels in the gap
between the $n$th and $(n+1)$th Landau bands \cite{Hal}.

\begin{figure}
\centerline{\psfig{figure=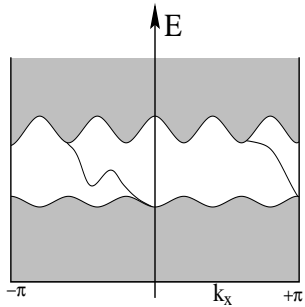,height=4.6cm,width=4.8cm,angle=0}}
\caption{{\sl 
Schematic representation
of the spectrum of $\hat{H}_H$ in a given gap of 
$H_H$, the solid lines are the Dirichlet bands and the shaded regions are the
Bloch bands. }} 
\end{figure}

\vspace{.2cm}

Now we would like to add a disordered potential $V$. First of all, if
$V$ is sufficiently small, sufficiently large gaps of $H_H$ remain
open for $H=H_H+V$. It follows further from Mourre estimates on the
current operator that the spectrum remains absolutely continuous in
the gaps of $H$ for a weak potential whenever the current of the edge
states of $\hat{H}_H$ has a definite  sign \cite{BP}. Whereas  the
latter condition is always satisfied for the Landau Hamiltonian, it
may not hold in the discrete case ({\sl cf.} Fig. 1 and the numerical
studies in \cite{Hat} where edge channels having edge states with
group velocity both to the left and to the right are exhibited).
In this situation, the positive commutator methods cannot be applied.
Nevertheless, we shall be able to show that the current remains 
constant. However, we cannot deduce that the spectrum is still 
absolutely continuous once a small perturbation is added. 

In order to treat the situation with broken translation invariance, we
parallel Bellissard's non-commutative generalization of the TKN$_2$
work \cite{Bel,BES}. No particular structure of the Hamiltonian $H$ on
$\ell^2(\ZZ^2)$ is needed  except for its homogenuity in the sense of
\cite{Bel,BES}.  The main mathematical tool in \cite{Bel,BES}  is the
C$^*$-algebra ${\cal A}$ of homogeneous observables in the plane. It
has the structure of a crossed product algebra ${\cal
A}=C(\Omega)\times{{\bf Z}_x}\times{{\bf Z}_y}$ associated to the
dynamical system given by the magnetic translations ${\bf Z}_x$ and
${\bf Z}_y$ in the $x$ and $y$-direction respectively acting on the
compact space of disorder configurations $\Omega$ which is the hull of
$H$.   Each such configuration $\omega\in\Omega$ induces a
representation $\pi_\omega$ of the observable algebra ${\cal A}$ on
physical Hilbert space $\ell^2({\bf Z}^2)$.  There exists an
$H\in{\cal A}$ such that $\pi_\omega(H)$ is precisely the Hamilton
operator with disordered configuration $\omega\in\Omega$.  
We now consider the Toeplitz extension $T({\cal A})$
with respect to the crossed product structure of ${\bf Z}_y$
\cite{PV}.  Its physical representations give operators in the half
plane. This naturally gives rise to an exact sequence of
$C^*$-algebras \cite{PV}
\begin{equation}
\label{eq-exact}
0\;\to\;{\cal E}
\;
\stackrel{{\scriptscriptstyle i}}{{\textstyle \to}}
\;
T({\cal A})
\;
\stackrel{{\scriptscriptstyle \pi}}{{\textstyle \to}}
\;
{\cal A}\;\to\;0
\mbox{ . }
\end{equation}
Here ${\cal E}$ is the C$^*$-algebra of observables localized near the
edge $y=0$; it is isomorphic to  the  C$^*$-tensor product of
$C(\Omega)\times{{\bf Z}_x}$ with the compact operators $K$.  The
exact sequence (\ref{eq-exact}) induces two six-term exact sequences,
one for K-theory groups \cite{PV,Bla} and one for the cyclic
cohomology groups \cite{Nes}, and we shall use their duality
\cite{Nes} to prove the equality of bulk and edge Hall conductivities.

Let us illustrate these notions for the Harper Hamiltonian with
arbitrary flux $\varphi$,  but without a further potential. The
C$^*$-algebra ${\cal A}$ is then the rotation algebra generated by the
two  magnetic translations $U_x$ and $U_y$ satisfying the commutation
relation $U_xU_y= e^{2\pi \imath \varphi}U_yU_x$. Thus in this case
$C(\Omega)\cong {\bf C}$. The Toeplitz extension is generated by
$\hat{U}_x$ and $\hat{U}_y$ satisfying the same commutation relation,
but, while $\hat{U}_x$ remains unitary, $\hat{U}_y$ is now only an
isometry satisfying $\hat{U}_y^*\hat{U}_y=1-\Pi_0$ where $\Pi_0$ is
the projection on the states supported by the boundary of $\Gamma$.
Finally, ${\cal E}$ is isomorphic to the tensor product of
C$^*(\hat{U}_x)\cong C(S^1)$ with $K$.  The maps in (\ref{eq-exact})
are the inclusion $i$ and the  projection $\pi$ given by
$\pi(\hat{U}_{x,y})=U_{x,y}$. 

The traces ${\cal T}_{x,y}$  of physical representations $\pi_\omega$
of an observable are almost surely independent of $\omega$ with
respect to any given invariant and ergodic measure ${\bf P}$ on
$\Omega$ \cite{BES}.  Hence they allow to define traces on the
observable algebras ${\cal A}$ and ${\cal E}$.  Now the definition
(\ref{eq-edgecond}) of the edge Hall conductivity   remains valid as
long as the projections $\hat{P}_J$ are in the Schatten ideal of
traceclass operators with respect to $\hat{{\cal T}}$ for $J$
sufficiently  close to $\{E\}$. This is possible even though
$\hat{P}_J$ is only an element of the bicommutant ${\cal E}''$, the
envelopping von  Neumann algebra.  Now the crucial observation is that
the current of the edge states in an interval $J$ lying in a gap $G$
of the spectrum of
$H$ can be calculated using Duhamel's formula and taking into account
elementary properties of projections:

\begin{equation}
\label{eq-Duh1}
\hat{{\cal T}}(\hat{P}_J \nabla_x(\hat{H}))
\;=\;
\frac{|J|}{2\pi \imath}
\;\hat{{\cal T}}((\hat{{\cal U}}(J)^*-1)\nabla_x\,\hat{{\cal U}}(J))
\mbox{ , }
\end{equation}
where
\begin{equation}
\label{eq-Duh2}
\hat{{\cal U}}(J)
\;=\;
\exp\left(2\pi\imath\, \hat{P}_J\frac{\hat{H}-E'}{|J|} \right)
\mbox{ , }
\qquad
E'=\inf(J)
\mbox{ . }
\end{equation}

\noindent Although $\hat{{\cal U}}(J)$ is built out of the operators
$\hat{P}_J$ and $\hat{H}$ which are not localized near the boundary
and not even in the C$^*$-algebra $T({\cal A})$, we can show that
$\hat{{\cal U}}(J)-1$ is an element of the edge algebra ${\cal E}$ by
using the exponential map  of the six-term exact sequence of
$K$-groups \cite{Bla}  associated to the exact sequence
(\ref{eq-exact}). More precisely,  the image under the exponential map
of the class $[P_\mu]_0\in K_0({\cal A})$  associated to the Fermi
projection $P_\mu$ is equal to the class $[\hat{{\cal U}}(J)]_1\in
K_1({\cal E})$  whenever the Fermi level $\mu$ is in $J$.   In fact,
$P_\mu$ is equal to the continuous function of the Hamiltonian
$f(H)=P_{E'}-P_J(H-E')/|J|$. Now a self-adjoint
lift of $P_\mu$ is given by $f(\hat{H})$. From
$[\hat{P}_{E'},\hat{P}_{J}]=0$ thus follows

\begin{equation}
\label{eq-exp}
\exp([P_\mu]_0) \;=\; [\exp(-2\pi\imath f(\hat{H}))]_1 \;=\;
[\hat{{\cal U}}(J)]_1 \mbox{ . }
\end{equation}

\noindent Finally we note that continuously varying the boundaries of 
$J$ to those of $G$ leads to a homotopy from $\hat{{\cal U}}(J)$ to
$\hat{{\cal U}}(G)$. Thus (\ref{eq-Duh2}) actually associates to $G$ 
a class in the K-group $K_1({\cal E})$.

It now follows from Connes' non-commutative geometry \cite{Con} that
$\frac{1}{\imath} \hat{{\cal T}}((\hat{{\cal
U}}^*-1)\nabla_x\,\hat{{\cal U}})$  is an integer for any  unitary
$\hat{{\cal U}}$ in (a suitable subalgebra of)  $\tilde{{\cal
E}}$. Actually $\zeta_1(\hat{A},\hat{B})=\frac{1}{\imath}
\hat{{\cal T}}(\hat{A}\nabla_x(\hat{B}))$ defines a 1-cocycle on
${\cal E}$  because $\hat{{\cal T}}$ is invariant under $\nabla_x$.
With some calculatory effort,  this cocycle can be linked to the
standard 1-cocyle of the Fredholm module $(C_1\otimes {\cal E}_0
,\pi_\omega\oplus\pi_\omega, \ell^2(\Gamma)\oplus \ell^2(\Gamma),
\sigma_2\otimes \imath X/|X| )$  where ${\cal E}_0$ is the in ${\cal
E}$ dense $*$-algebra of operators with finite support in
$y$-direction and $C_1$ a two-dimensional ${\bf Z}_2$ graded Clifford
algebra in $\mbox{Mat}({\bf C}^2)$, $\pi_\omega\oplus\pi_\omega$ is a
doubling of the physical representation  on the doubled physical
Hilbert space $\ell^2(\Gamma)\oplus \ell^2(\Gamma)$  and the Dirac
phase is constructed from the Pauli matrix $\sigma_2$ and the position
operator $X$.  Hence the odd index theorem \cite[p. 291]{Con}, a
density and homotopy argument linking 
$\hat{{\cal U}}(G)$ to an element in ${\cal E}_0$ \cite[p. 249]{Con}
and a
treatment of the disorder configuration along the lines of \cite{BES}
imply the following result.

\vspace{.2cm}

\noindent {\bf Theorem 1} 
{\sl Suppose that $G\subset{\bf R}$ is a spectral gap of the plane
operator $H$ acting on $\ell^2(\ZZ^2)$. Let $\Pi$
denote the projection from $\ell^2({\bf Z}\otimes{\bf N})$ onto
$\ell^2({\bf N}\otimes{\bf N})$ and let $\hat{{\cal U}}(G)$ 
constructed by {\rm (\ref{eq-Duh2})} from $\hat{P}_G$.  Then for 
${\bf P}$-almost every $\omega\in\Omega$, the
operator $\Pi \pi_\omega(\hat{{\cal U}}(G))\Pi$ is a Fredholm operator
on $\ell^2({\bf N}\otimes{\bf N})$ with constant index 
and for all $E\in G$

$$
\sigma^e_{\perp}(E)
\;=\;
\frac{q^2}{h}\,
\mbox{\rm Ind}
\left(\Pi\,\pi_\omega(\hat{{\cal U}}(G))\,\Pi
\right)
\mbox{ . }
$$
}

\vspace{.2cm}

We remark that the index can also be written as a relative index of
a pair of projections as defined by Avron, Seiler and Simon \cite{ASS},
notably as the relative index of $\Pi$ and
$\pi_\omega(\hat{{\cal U}}(G))\Pi\pi_\omega(\hat{{\cal U}}(G))^*$.

\vspace{.2cm}

Using the exact sequence (\ref{eq-exact}), we now link this edge
theory to the bulk theory as developed in \cite{BES}.  From the above
follows that $\sigma^e_{\perp}(E)$ actually results from a pairing 
\cite{Con}
between  $[\hat{{\cal U}}(G)]_1\in K_1({\cal E})$ with the odd cyclic
cohomology  class defined by the 1-cocycle $\zeta_1$ given above.
Similarly, the bulk Hall conductivity $\sigma^b_{\perp}(\mu)$ for a
Fermi level $\mu$ in a gap of $H$ comes from a pairing of the class
of the Fermi projection $[P_\mu]_0\in K_0({\cal A})$ with the
2-cocycle $\zeta_2$ over ${\cal A}$ defined by $\zeta_2(A,B,C)=2\pi
\imath \TV_x\TV_y(A\nabla_x B \nabla_y C- A\nabla_y B \nabla_x C)$
\cite{BES}: 
$$
\sigma^e_{\perp}(E)
\;=\;
\langle \zeta_1,[\hat{{\cal U}}(G)]_1\rangle
\mbox{ , }
\qquad
\sigma^b_{\perp}(\mu)
\;=\;
\langle \zeta_2,[P_\mu]_0\rangle
\mbox{ . }
$$
We showed above that $[\hat{{\cal U}}(G)]_1$ is the image of
$[P_\mu]_0$ under the
exponential map of K-theory.
Next one can verify that 
the 1-cocycle $\zeta_1$ over ${\cal E}$ is mapped to the 2-cocyle
$\zeta_2$ over ${\cal A}$ under the mapping \# defined in
\cite[Sec. 8]{Nes}.  For this map, the duality theorem of the pairing
holds, notably $\langle \zeta_1,\exp([P]_0)\rangle =\langle
\#\zeta_1,[P]_0\rangle$ for any projection $P\in\Aa$
\cite[Sec. 12]{Nes}. Hence we obtain:

\vspace{.2cm}

\noindent {\bf Theorem 2} 
{\sl 
$\sigma^e_{\perp}(E)=\sigma^b_{\perp}(E)$  for all
energies $E$ in a spectral gap of $H$.  }

\vspace{.2cm}

At this point, let us comment on generalizations of these results.  Just
as one does not need the existence of a gap $G$ in order to prove the
quantization of the bulk Hall conductivity \cite{BES,AG}, it is
likely that Theorems 1 and 2 hold under the
weaker hypothesis that the interval $G$ only contains dynamically
localized states of $H$ in the sense of \cite{BES}.  Furthermore, the
whole theory should have a continuous analogon for a disordered Landau
Hamiltonian.  As both of these results ask for more lengthy
and detailed proofs, they will be subject of a forthcoming
publication.  

\vspace{.2cm}

We now sketch how the above results lead to the desired explanation of
a QH regime measurement in a QH bar.  Following Halperin \cite{Hal},
we suppose that the measured Hall voltage $V_\perp$ across the bar is
the sum of the the potential drop $V^b$ due to an electrostatic field
and the (relativ)  chemical potential difference $\Delta\mu/q=
(\mu_{{\tiny\rm u}}-\mu_{{\tiny\rm l}})/q$  between the upper and the
lower edge.  Furthermore, let the interval $[\mu_{{\tiny\rm l}},
\mu_{{\tiny\rm u}}]$ be contained in a gap $G$ of $H$ (the above
generalization only needs the weaker condition that $G$ is dynamically
localized). In linear response approximation, the  electric field
leads to a bulk current $I^b=\sigma^b_{\perp} V^b$. Now both the upper
and the lower edge may carry a current.  In absence of backscattering,
we can treat them as two separate half-plane problems. But actually
the lower edge can be seen as an upper edge with reversed  magnetic
field, which is equivalent to a time reversal. This changes the
orientation of its current so that the net current carried by both
edges comes from  the upper edge states with energies in
$[\mu_{{\tiny\rm l}},\mu_{{\tiny\rm u}}]$. From the above thus results
a net edge current $I^e=\sigma^e_{\perp}\Delta\mu /q$.  Hence the Hall
conductance of the sample given by the quotient of the total current
$I=I^e+I^b$ and the voltage $V_\perp$  is equal to the integer
$\sigma^e_{\perp}=\sigma^b_{\perp}$  for any value of $V^b/V_\perp$.

\vspace{.2cm}

An interesting question which has led to considerable theoretical and
experimental work (see \cite{YTP} and references therein) is how much
current is carried by either edge or bulk states in a typical QH
experiment.   Let us argue that at most 10\% of the current is carried
by the edge states. This agrees with recent experimental studies
\cite{YTP}.  For the edge current of $[\mu_{{\tiny\rm
l}},\mu_{{\tiny\rm u}}]$ to be equal to an integer times $\Delta\mu$,
the difference of chemical potentials $\Delta \mu$ clearly has to be
smaller than the energetic distance $\hbar\omega_c(1-p)$ (here
$\omega_c$ is the cyclotron frequency so that $\hbar\omega_c$ is the
distance between two Landau levels, and $p$ is the quotient of the
energetic width of the plateaux and $\hbar\omega_c$). Hence the
proportion of edge current has to be smaller than $\Delta
\mu/pV_\perp$.  In order to estimate this condition and the
temperature corrections below, we use the experimental data from
\cite[Chapter 2]{PG} for the $\sigma_\perp=4$ plateau: $B\approx
6\,T$, $V_\perp\approx 170\,mV$ and $T\approx 1.2\, K$ and $p\approx
0.6$. Using the data for the effective electron mass ($m_*\approx
0.07m_e$) and the electron charge, we obtain $\hbar\omega_c\approx
48\, meV$ and a maximal proportion of edge currents of 10\%. 

\vspace{.2cm}

We acknowledge support by the SFB 288.


\end{document}